\documentclass[prb,amsmath,showpacs,twocolumn,floatfix]{revtex4-1}

\usepackage{amssymb}
\usepackage{graphicx}
\usepackage{subfigure}
\usepackage{color}
\usepackage{multirow}
\usepackage{verbatim}
\usepackage{array}

\begin{document}

%%%%%%%%%%%%%%%%%%%%%%%%%%%%%%%%%%%%%%%%%%%%%%%%%%%%%%%%%%%%%%%%%%%%%%%%%%%%%%%%

\title{
Epitaxial Phases of BiMnO$_3$ from First Principles
}

\author{Oswaldo Di\'eguez$^1$ and Jorge \'I\~niguez$^{2,3}$}

\affiliation{\vspace{0.3cm}
             $^1$Department of Materials Science and Engineering, 
             Faculty of Engineering, Tel Aviv University,
             Tel Aviv 69978, Israel 
             \vspace{0.3cm} \\
             $^2$Materials Research and Technology Department, Luxembourg
             Institute of Science and Technology, 5 avenue des Hauts-Fourneaux,
             L-4362 Esch/Alzette, Luxembourg 
             \vspace{0.3cm} \\
             $^3$Institut de Ci\`encia de Materials de Barcelona (ICMAB-CSIC),
             Campus UAB, 08193 Bellaterra, Spain
             \vspace{0.3cm} \\}

%%%%%%%%%%%%%%%%%%%%%%%%%%%%%%%%%%%%%%%%%%%%%%%%%%%%%%%%%%%%%%%%%%%%%%%%%%%%%%%%

\begin{abstract}
Bulk BiMnO$_3$ is the only transition-metal perovskite oxide that is
insulating and shows strong ferromagnetism. 
This distinctive behavior
would make it a promising candidate as a magnetoelectric multiferroic
if it was also a polar material, but experiments have shown that
bulk BiMnO$_3$ has either a very small polarization (below
0.1~$\mu$C/cm$^2$) or, most likely, that it is a paraelectric.
There is also experimental evidence that the
polarization in BiMnO$_3$ {\em films} grown on SrTiO$_3$ can be as
high as 20~$\mu$C/cm$^2$.
Despite of the interest of these behaviors, the
diagram of BiMnO$_3$ as a function of epitaxial strain has remained
largely unexplored.
In this article, we use first-principles to predict that both under
enough compressive and tensile epitaxial strain BiMnO$_3$ films are
ferroelectric with a giant polarization around 100~$\mu$C/cm$^2$. The
phases displayed by the films are similar to those experimentally
found for BiFeO$_3$ in similar conditions---at compressive strains,
the film is supertetragonal with a large component of the polarization
pointing out of plane, while at tensile strains the polarization
points mostly in plane. 
Like in BiFeO$_3$ films, these phases are antiferromagnetic---the
orbital ordering responsible for ferromagnetism in BiMnO$_3$ is absent
in the polar phases.
Our calculations also show that the band gap
of some of these BiMnO$_3$ films is substantially smaller than gaps
typically found in ferroelectric oxides, suggesting it may be a
suitable material for photovoltaic applications.
\end{abstract}

\date{\today}


% 77. Dielectrics, piezoelectrics, and ferroelectrics and their properties
% 77.80.-e Ferroelectricity and antiferroelectricity
% 77.84.-s Dielectric, piezoelectric, ferroelectric, and antiferroelectric materials

% 75. Magnetic properties and materials
% 75.80.+q Magnetomechanical and magnetoelectric effects,
% magnetostriction
% 75.85.+t Magnetoelectric effects, multiferroics

% 71.  Electronic structure of bulk materials
% 71.15.Mb Density functional theory, local density approximation,
% gradient and other corrections 

\maketitle

%%%%%%%%%%%%%%%%%%%%%%%%%%%%%%%%%%%%%%%%%%%%%%%%%%%%%%%%%%%%%%%%%%%%%%%%%%%%%%%%

\section{Introduction}

BiMnO$_3$ is the only strong ferromagnetic insulator among the transition-metal
perovskite oxides, a family of functional materials whose members
display many different properties of technological interest.
The Curie temperature of BiMnO$_3$ is around
105~K,\cite{Bokov1966SPSS, Sugawara1968JPSJ} below which 
several groups have reported measurements of its magnetic moment close to
4~$\mu_{\rm B}$ per Mn atom.\cite{Moreira2002PRB, Kimura2003PRB}
The unique ferromagnetic behaviour of BiMnO$_3$ 
is related to the presence of orbital ordering
of the Mn$^{+3}$ ions (3$d^4$)---elongated half-filled $d_z^2$ orbitals
point towards empty $d_{x^2-y^2}$ orbitals of neighboring Mn
cations.\cite{Moreira2002PRB}
For many years BiMnO$_3$ was supposed to have also a weak switchable
polarization (below 0.1~$\mu C$/cm$^2$), and the polar space group $C2$ was
assigned to its crystal
structure.\cite{Atou1999JSSC, Moreira2002PRB}
However, recent electron diffraction experiments by Belik and 
coworkers\cite{Belik2007JACS} point to a $C2/c$ space group, which is
not polar; these results have been
confirmed by neutron powder diffraction experiments.\cite{Montanari2007PRB}
Despite of this, when BiMnO$_3$ is grown as a
epitaxial film on SrTiO$_3$ it develops a strong polarization, which Jeen
and coworkers\cite{Jeen2011JAP} measured to be
23~$\mu C$/cm$^2$; other
groups reported similar results---a polarization between 9 and 
16~$\mu C$/cm$^2$,\cite{Son2008APL} and signatures of
ferroelectricity.\cite{DeLuca2013APL}

Perovskite oxides display dramatic changes in their properties
when grown as epitaxial films on a substrate, with
the misfit strain imposed by the substrate 
playing a major role in these changes.\cite{Schlom2007ARMR}
Given that other nonpolar perovskites 
develop ferroelectricity under epitaxial strain, 
it is somewhat surprising that the epitaxial phase diagram of
BiMnO$_3$ is largely unexplored in the search for possible coexistence
of magnetic and polar orderings.
Using first-principles calculations, Hatt and Spaldin\cite{Hatt2009EPJB}
concluded that BiMnO$_3$ remains non-polar when compressive or tensile strain
is applied to the film; they used the $C2/c$ bulk phase
as the starting point of their calculations, subjecting the simulation cell to 
constraints that mimic the strained film.
No other phases seem to have been studied using computational
approaches, although earlier calculations by Hill and Rabe\cite{Hill1999PRB}
and by Seshadri and Hill\cite{Seshadri2001CM} pointed to the
existence of ferroelectric instabilities in this material.
In this study we explore structures that are good candidates
to become the ground state of epitaxial BiMnO$_3$ films.
At high compressive strains, we have looked at 
the supertetragonal phases with giant polarization that are found in 
BiFeO$_3$ films\cite{Bea2009PRL} and in BiCoO$_3$ bulk and
films;\cite{Belik2006CM} these phases are called supertetragonal, or $T$,
because they have a large $c/a$ ratio (some of them have lower
symmetry than tetragonal).
At high tensile strains, we have explored phases of low energy
in BiFeO$_3$ (including the $Pnma$ structure of both bulk BiFeO$_3$ and
bulk BiMnO$_3$ under pressure\cite{Guennou2014PRL}
and epitaxial polar phases of BiFeO$_3$ under tensile strains).
We have found that BiMnO$_3$ and BiFeO$_3$ films display similar structures
under high tensile and compressive strains; therefore, we predict that
BiMnO$_3$ has a strong polarization when grown in these
conditions.

%%%%%%%%%%%%%%%%%%%%%%%%%%%%%%%%%%%%%%%%%%%%%%%%%%%%%%%%%%%%%%%%%%%%%%%%%%%%%%%%

\section{Methods}

Our first-principles calculations are based on density-functional theory
(DFT).\cite{Hohenberg1964PR,Kohn1965PR}
However, BiMnO$_3$ contains highly localized $d$
orbitals that are not described accurately within pure DFT.
One approach towards a better description of the electronic properties of these
solids is to add a ``Hubbard $U$'' term to the energy of the system that favors
localization on those electrons; this requires picking
a value of $U$ that reproduces some set of experimental or more accurate
theory results.
Another approach
is to use a hybrid density functional that includes a portion of exact
Hartree-Fock exchange;\cite{Becke1993JCP} this in general ameliorates the
problems with $d$ or $f$ electron delocalization, predicts band gaps
for solids that are much closer to experimental results than those of pure
DFT,\cite{Krakau2006JCP} and performs better than the
``Hubbard $U$'' approach 
in perovskite oxides such as BiFeO$_3$.\cite{Stroppa2010PCCP}
Both approaches are implemented in {\sc Vasp},\cite{VASP} the 
first-principles code that we have used to carry out the calculations
presented in this work.
To optimize our typical structures with {\sc Vasp} the second approach
requires a hundred times more computer time than the
first one, so we have used the code in the following way:
for our exploratory calculations we applied a Hubbard $U$ following the 
rotationally invariant method described in
Ref.~\onlinecite{Liechtenstein1995PRB},
and for our final results we used a hybrid functional according to the HSE06 
prescription.\cite{Krakau2006JCP}
As in our previous article on BiFeO$_3$,\cite{Dieguez2011PRB} we 
worked with the Perdew-Burke-Ernzerhof DFT exchange-correlation functional 
adapted to solids (PBEsol).\cite{Perdew2008PRL}
We used the projector augmented-wave method to represent the ionic
cores,\cite{Blochl1994PRB}
solving for the following electrons: Mn's $3p$, $3d$, and 
$4s$; Bi's $5d$, $6s$, and $6p$; and O's $2s$ and $2p$.
We represented wave functions in a plane-wave basis set truncated at 500 eV.
We performed integrations within the Brillouin zone by using $k$-point grids
with densities similar to that of a
$6 \times 6 \times 6$ grid for a 5-atom perovskite
unit cell.

%%%%%%%%%%%%%%%%%%%%%%%%%%%%%%%%%%%%%%%%%%%%%%%%%%%%%%%%%%%%%%%%%%%%%%%%%%%%%%%%

\section{Results and Discussion}

\subsection{Metastable Phases of Bulk BiMnO$_3$}

We started our search for metastable phases of BiMnO$_3$ by doing
PBEsol+$U$ ionic relaxations for the seven lowest-energy crystal structures 
found for bulk BiFeO$_3$ in our previous work\cite{Dieguez2011PRB}
(which can accommodate to a $2 \times 2 \times 2$ pseudocubic 40-atom
cell) and for
the experimental $C2/c$ ground state configuration of BiMnO$_3$
(using 40-atom cells that are not pseudocubic).
For each of these eight configurations we prepared four types of magnetic
arrangments, as in Ref.~\onlinecite{Dieguez2011PRB}---ferromagnetic (FM), and
antiferromagnetic of the A, C, and G types (A-AFM, C-AFM, G-AFM, respectively).
We then did three types of searches for local minima of the
energy: (1) we directly relaxed the structures until forces and
stresses were close to zero; (2) we did a few steps of molecular dynamics
in order to break possible spureous symmetries, and then we relaxed the
resulting structures until forces and stresses were close to zero; and (3) we 
took the lowest-energy magnetic ordering found so far for each structure type
and relaxed the atoms imposing each of the other three magnetic orderings.
At the end, we chose the lowest-energy phase for each type of structure found
and for each type of magnetic ordering.
In all the optimization calculations of this work
the final forces were below 0.015~eV/\AA~and the final stresses are below
0.0005~eV/\AA$^3$.

\begin{figure}
\centering
\includegraphics[width=75mm]{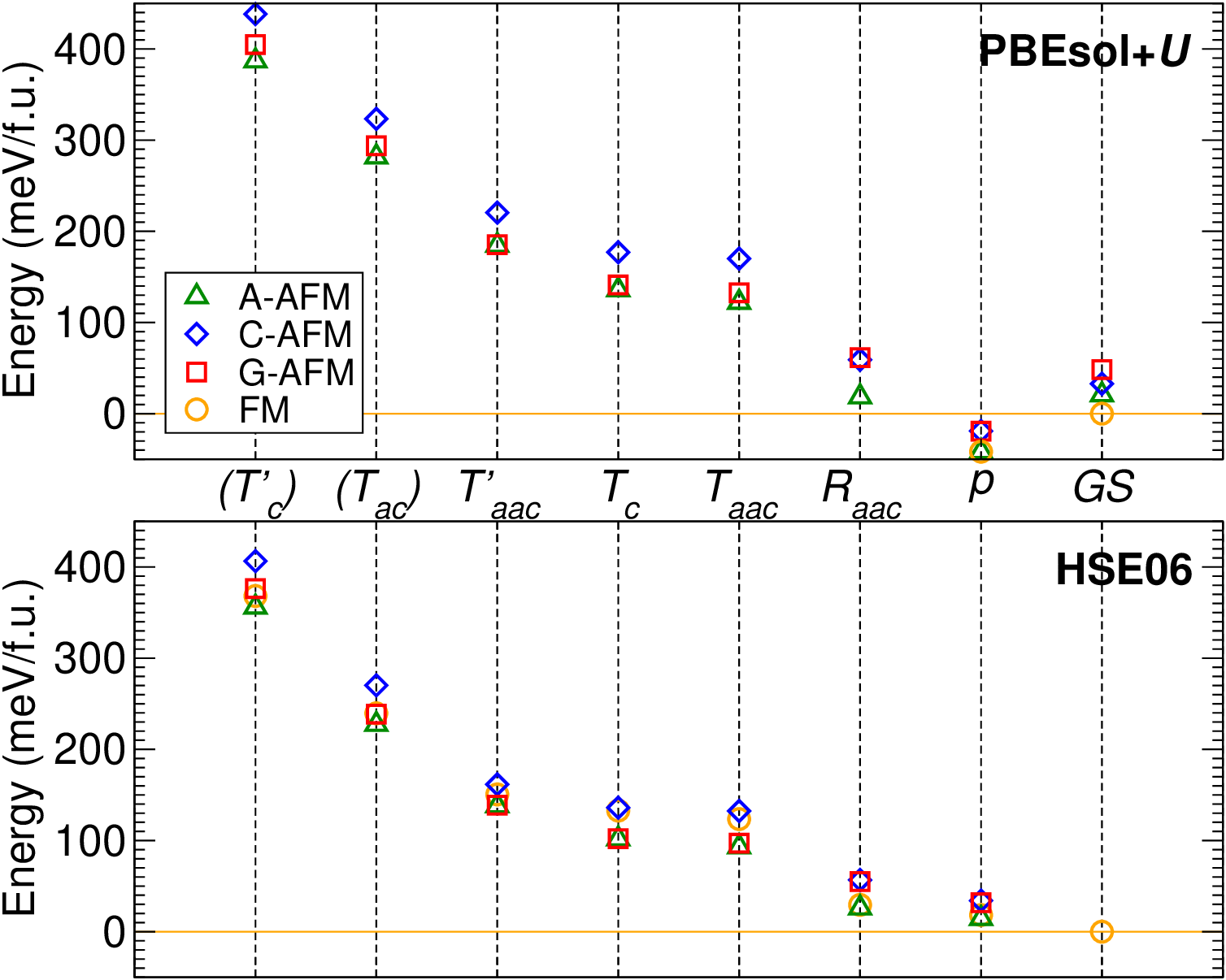}
\caption{(Color online.) 
Energies of the bulk BiMnO$_3$ phases found to be either 
saddle points of the energy surface ($T_c'$ and $T_{ac}$)
or local minima (the rest),
computed using PBEsol+$U$ (top) and HSE06 (bottom), for each of four
magnetic orderings.
}
\label{fig_energies}
\end{figure}

The process just described lead to the identification of the
configurations whose energies are given in Fig.~\ref{fig_energies} (top).
The corresponding 
phases are labeled in the following way
(directions are given in the pseudocubic setting): 
the ground state with $C2/c$
symmetry is labeled as GS; the paraelectric phase with
$Pnma$ symmetry is labeled as $p$; the ferroelectric phase derived from the
$R3c$ phase that is the ground state of BiFeO$_3$ is labeled as $R_{aac}$
(since Mn$^{+3}$ is a $d^4$ Jahn-Teller active ion, 
the original $R3c$ phase distorts into this one, which has
a polarization with a component along $[110]$
and another one along $[001]$, corresponding to the $Cc$ monoclinic 
space group);
the other phases are
supertetragonal $T$ phases like the ones mentioned earlier.
Three of these $T$ phases are local minima of the energy
according to our analysis of the corresponding force-constant matrices:
$T_{aac}$ (originating from the $Cc$ phase, with a small component of the
polarization along $[110]$ and a large one along $[001]$),
$T_c$ (originating from the $Pna2_1$ phase, with polarization along $[001]$),
and
$T_{aac}'$ (originating from the $Pc$ phase, with a small component of the
polarization along $[110]$ and a large one along $[001]$).
We also found that two of these $T$ phases are {\em not} local minima of the
energy:
$T_{ac}$ (originating from the $Cm$ phase, with a small component of the
polarization along $[010]$ and a large one along $[001]$) 
and
$T_c'$ (the simplest $P4mm$ tetragonal configuration with
polarization along $[001]$).

The calculations mentioned in the previous paragraph were done using
$U = 4$~eV and $J = 1$~eV, since these values gave good agreement between the
structural parameters computed for the bulk ground state and the experimental
results; however, the obtained band gap was close to zero, and reasonable
variations of $U$
and $J$ could not get the band gap to open above 0.5~eV (the 
experimental gap of FM BiMnO$_3$ is 0.9 eV,\cite{McLeod2010PRB} but it is
expected to be larger for antiferromagnetic configurations).
Because of this failure to open band gaps, in many cases when optimizing FM
structures we would get a spureous metallic state; this is the reason why
in Fig.~\ref{fig_energies} (top) many of the FM states are absent.
Another spureous result obtained when using PBEsol+$U$ is that the experimental
ground state of BiMnO$_3$ is predicted to have higher energy than the $p$
phases.

We then used the more computationally demanding HSE06 hybrid
method to re-optimize
the structures reported in Fig.~\ref{fig_energies} (top), and similar FM
structures; for all structures but some of the GS ones it is possible to do
this with 20-atom unit cells.
All phases were found to be robust insulators, and the $C2/c$ (GS) structure
was accurately predicted as the one with the lowest energy.
Otherwise, the energy differences between most phases
are similar to those obtained using PBEsol+$U$.
This is in line with the previous finding for BiFeO$_3$ in which we reported
that energy differences between {\em similar} structures are independent
of the exchange-correlation functional used due to
error cancelation.\cite{Dieguez2011PRB}
In particular, PBEsol+$U$ and HSE06 predict very similar energy differences
between different magnetic arrangements of the same structure; this is why
we have not recomputed other magnetic orderings of the GS phase (some of 
them require to double the unit cell, making these calculations computationally
expensive).
For each of the geometries that are local minima of the energy,
Tables~\ref{tab_bulk} and \ref{tab_bulkwickoff}
list in detail the properties of the structure having the 
favored magnetic ordering; in particular, the lattice parameters predicted
for the $C2/c$ phase are within 0.4\% of the experimental 
values.\cite{Belik2007JACS}
The $T$ phases have similar characteristics
to those found for BiFeO$_3$ (an analysis of most of their atomic
displacement patterns can be found in Ref.~\onlinecite{Dieguez2011PRB});
the most stable one has also a $Cc$ space group,
a large $c/a$ ratio (1.17),
and a large macroscopic polarization (94~$\mu$C/cm$^2$).
Magnetic arrangements other than those shown in these Tables display similar 
structural configurations, differences in their lattice parameters being
always below 1\%.
The band gaps are always smallest for the FM ordering, and
largest for the G-AFM orderings, the difference for a given structure
going up to 1.5~eV.

\begin{table}
\setlength{\extrarowheight}{0.5mm}
\setlength{\tabcolsep}{5pt}
\caption{Properties of the bulk BiMnO$_3$ phases 
corresponding to minima of the energy found in this work:
name given to the phase (and most favorable magnetic
ordering), space group, energy (in meV per formula unit with respect to the
ground state), band gap (in eV), $c/a$ ratio, 
and polarization vector (in $\mu$C/cm$^2$). 
}
\begin{tabular}{ccccc}
\hline\hline
Phase & S.G. &  $\Delta E$ & Gap & ${\bf P}$ \\
\hline
$T_{aac}'$  (A-AFM)  &  $Pc$      &  138  &  2.1  &    $(42, 42, 95)$  \\
$T_{c}   $  (A-AFM)  &  $Pna2_1$  &  102  &  1.8  &    $(0,0,75)$  \\
$T_{aac} $  (A-AFM)  &  $Cc$      &   93  &  1.8  &    $(43, 43, 73)$  \\
$R_{aac} $  (A-AFM)  &  $Cc$      &   26  &  2.7  &    $(61, 61, 39)$  \\
$p       $  (A-AFM)  &  $Pnma$    &   15  &  2.7  &    $(0,0,0)$  \\
SG          (FM)     &  $C2/c$    &    0  &  1.7  &    $(0,0,0)$  \\
\hline\hline
\end{tabular}
\label{tab_bulk}
\end{table}

\begin{table}
\setlength{\extrarowheight}{0.5mm}
\setlength{\tabcolsep}{5pt}
\caption{Lattice parameters and Wickoff positions of the bulk BiMnO$_3$
phases found to be a minimum of the energy in this work.}
\vskip 0mm
\begin{tabular}{cccccc}
\hline\hline
Phase & \multicolumn{5}{c}{Structure} \\
\hline
$T_{aac}'$
 &
\multicolumn{5}{c}{
$a=4.597$~\AA, $b=5.230$~\AA, $c=5.256$~\AA
}\\
 &
\multicolumn{5}{c}{
$\alpha=90^\circ$, $\beta=91.3^\circ$, $\gamma=90^\circ$
}\\
               &   Mn  &  2a & 0.5667 & 0.2552  & -0.0557 \\
               &   Bi  &  2a & 0      & 0.7799  & 0       \\
               &   O   &  2a & 0.1270 & 0.8051  & 0.4122  \\
               &   O   &  2a & 0.6226 & 0.5461  & 0.1486  \\
               &   O   &  2a & 0.6709 & -0.0276 & 0.7141  \\
\hline
$T_a$
 &
\multicolumn{5}{c}{
$a=5.315$~\AA, $b=5.214$~\AA, $c=8.764$~\AA
}\\
 &
\multicolumn{5}{c}{
$\alpha=90^\circ$, $\beta=90^\circ$, $\gamma=90^\circ$
}\\
               &   Mn  &  4a & 0       & -0.0046 &  0.2788  \\
               &   Bi  &  4a & 0.5556  & -0.0111 &  0       \\
               &   O   &  4a & -0.0303 & -0.0750 &  0.0429  \\
               &   O   &  4a & 0.8105  &  0.3081 &  0.2885  \\
               &   O   &  4a & 0.7120  &  0.7921 &  0.3282  \\
\hline
$T_{aac}$
 &
\multicolumn{5}{c}{
$a=10.235$~\AA, $b=5.233$~\AA, $c=5.310$~\AA
}\\
 &
\multicolumn{5}{c}{
$\alpha=90^\circ$, $\beta=121.2^\circ$, $\gamma=90^\circ$
}\\
               &   Mn  &  4a & 0.2210 & 0.2546 & 0.1668 \\
               &   Bi  &  4a & 1/2    & 0.2602 & 0      \\
               &   O   &  4a & 0.4572 & 0.3218 & 0.3688 \\
               &   O   &  4a & 0.7124 & 0.4440 & 0.3471 \\
               &   O   &  4a & 0.1703 & 0.4634 & 0.3953 \\
\hline
$R_{aac}$
 &
\multicolumn{5}{c}{
$a=9.266$~\AA, $b=5.720$~\AA, $c=5.675$~\AA
}\\
 &
\multicolumn{5}{c}{
$\alpha=90^\circ$, $\beta=125.7^\circ$, $\gamma=90^\circ$
}\\
               &   Mn  &  4a & 0.5233  & 0.2458 & 0.3328 \\
               &   Bi  &  4a & 0.2422  & 0.2422 & 0.4844 \\
               &   O   &  4a & 0.2744  & 0.3161 & 0.1236 \\
               &   O   &  4a & -0.0039 & 0.4427 & 0.0890 \\
               &   O   &  4a & 0.5754  & 0.4725 & 0.1539 \\
\hline
$p$
 &
\multicolumn{5}{c}{
$a=5.928$~\AA, $b=7.440$~\AA, $c=5.372$~\AA
}\\
 &
\multicolumn{5}{c}{
$\alpha=90^\circ$, $\beta=90^\circ$, $\gamma=90^\circ$
}\\
               &   Mn  &  4b & 0        & 0       & 1/2     \\
               &   Bi  &  4c & 0.07245  & 1/4     & 0.01010 \\
               &   O   &  4c & -0.02765 & 1/4     & 0.59634 \\
               &   O   &  8d & 0.82708  & 0.45595 & 0.20851 \\
\hline
GS
 &
\multicolumn{5}{c}{
$a=9.544$~\AA, $b=5.592$~\AA, $c=9.852$~\AA
}\\
 &
\multicolumn{5}{c}{
$\alpha=90^\circ$, $\beta=110.8^\circ$, $\gamma=90^\circ$
}\\
               &   Mn  &  4e & 0      & 0.2921 & 1/4    \\
               &   Mn  &  4c & 1/4    & 1/4    & 0      \\
               &   Bi  &  8f & 0.6353 & 0.2270 & 0.1215 \\
               &   O   &  8f & 0.5977 & 0.1733 & 0.5804 \\
               &   O   &  8f & 0.1455 & 0.0718 & 0.3737 \\
               &   O   &  8f & 0.3518 & 0.0472 & 0.1652 \\
\hline\hline
\end{tabular}
\label{tab_bulkwickoff}
\end{table}

\subsection{Structures of BiMnO$_3$ Epitaxial Films}

\begin{figure}
\centering
\includegraphics[width=75mm]{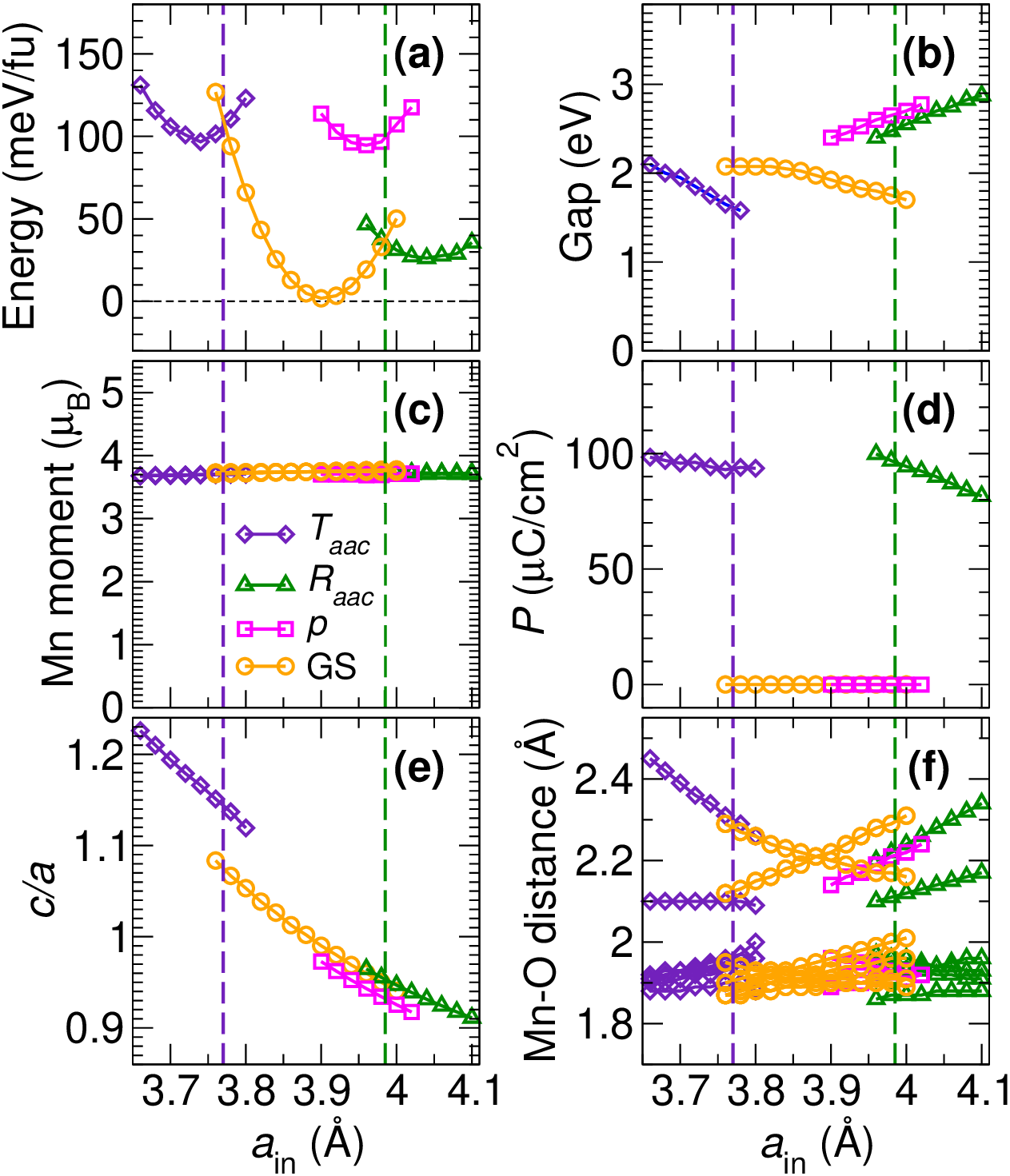}
\caption{(Color online.) Properties of BiMnO$_3$ films as a function of the
in-plane lattice parameter:
(a) energy relative to the bulk ground state;
(b) band gap;
(c) magnetic moment of the Mn ions;
(d) macroscopic polarization;
(e) $c/a$ ratio;
and (f) Mn--O distances.
The four structures mentioned in the legend of (c) have been considered in 
their favored magnetic ordering (FM for GS and A-AFM for the rest).
}
\label{fig_films}
\end{figure}

The bulk phases described here can be made stable by
growing the material as a coherent epitaxial thin film.
Like in BiFeO$_3,$\cite{Dieguez2011PRB} the supertetragonal phases are 
expected to be favored at compressive strains; at tensile strains, the
large in-plane lattice parameters of the $p$ and $R_{aac}$ phases
hint that these might be more stable than the GS phase.
To check these hypothesis we have done structural optimizations for the four
lowest-energy phases, where the epitaxial effect
is simulated by constraining the in-plane lattice vectors to be
equal in length and to form a 90$^\circ$ angle.
Figure~\ref{fig_films} shows how the properties of these films change
with the in-plane lattice constant.
At lattice constants around 3.90~\AA\ the epitaxial distortion on the GS film
breaks the $C2/c$ symmetry, but it is otherwise very small, so the 
corresponding film is still the favored one.
As we compress the film, the $T_{aac}$ phase becomes competitive, and below
around
3.75~\AA\ it is expected to be the stable state of the material.
For tensile strains at in-plane lattice parameters
above 4~\AA\ we expect the $R_{aac}$ film to be the ground state, displaying
a large polarization with components both in plane and out of plane.
The $p$ phase is, by far, the one that has to be most distorted to fit the
square symmetry of the substrate, and this renders it energetically not 
competitive with the other ones. 
The films are always insulating and display
magnetic moments of around 4~$\mu_{\rm B}$ localized in the Mn ions.
Their $c/a$ ratios grow markedly as the strain becomes more compressive, and
this translates partially into larger out-of-plane polarizations.
Most of the Mn--O bonds stay at values of around 1.9~\AA, but we can also see
in Fig.~\ref{fig_films} much longer bonds; those arise due to 
the orbital ordering in the GS phase, and in the out-of-plane directions
of the $T$ phases.

\subsection{Magnetic Properties}

\begin{figure}
\centering
\subfigure[]{
\includegraphics[width=75mm]{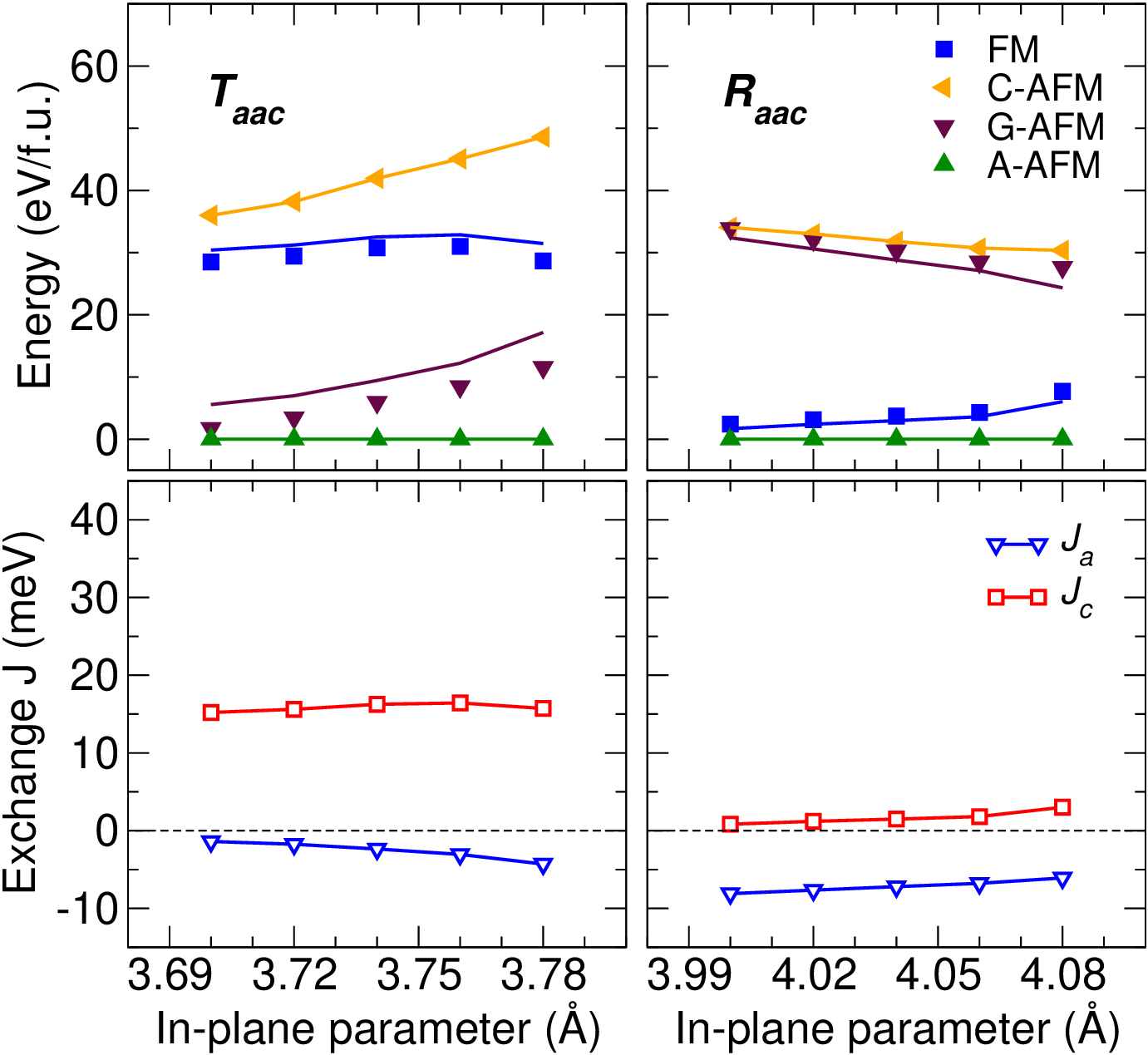}
}
\\
\subfigure[]{
\includegraphics[width=75mm]{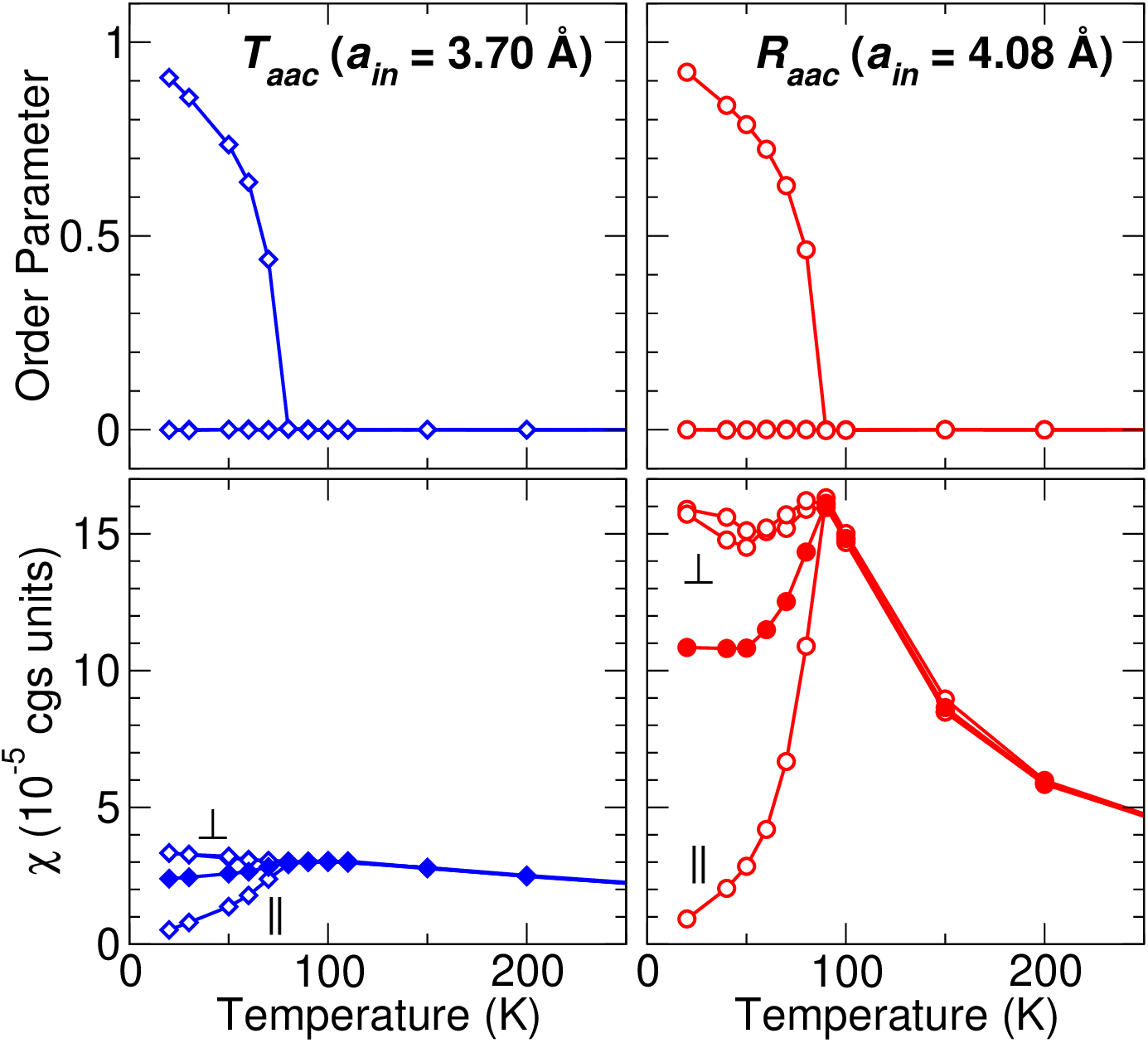}
}
\caption{(Color online.)
(a)
Top panels: energy of different magnetic arrangements with respect to the
most stable one (A-AFM), for the $T_{aac}$ (left) and $R_{aac}$ (right) phases.
Bottom panels: exchange constants $J$ from fitting 
those energies to a simple Heisenberg model, for the 
$T_{aac}$ (left) and $R_{aac}$ (right) phases.
(b) 
Top panels: A-AFM order parameter as a function of temperature for a $T_{aac}$
film of in-plane lattice constant of 3.70~\AA\ (left) and for a $R_{aac}$ 
film of in-plane lattice constant of 4.08~\AA\ (right).
Bottom panels: total magnetic susceptibilities for the same films (filled
symbols) and their partial contributions, parallel ($\parallel$)
and perpendicular
($\perp$) to the direction of the A-AFM order parameter (empty symbols). 
}
\label{fig_magnetism}

\end{figure}

The different magnetic orderings of a particular BiMnO$_3$ structure are within
around 50~meV/f.u., both in bulk (Fig.~\ref{fig_energies})
and in films (Fig.~\ref{fig_magnetism}(a), top panels).
The A-AFM ordering is favored both for the $T_{acc}$ and for the $R_{aac}$ 
phases.
These phases cannot accommodate the orbital ordering of the bulk, so the 
ferromagnetic ordering is not the prefered one any more.

To estimate the N\'eel temperature for the $T_{aac}$ and $R_{aac}$ phases
we used a Heisenberg model with energy
$E = E_0 + 1/2 \sum_{ij} J_{ij} {\bf S}_i \cdot {\bf S}_j$; $E_0$ is a 
reference energy, and $J_{ij}$ is the exchange coupling constant between the
spins localized at Mn ions $i$ and $j$, given by ${\bf S}_i$ and ${\bf S}_j$
(taken as unit vectors).
We restrict ourselves to first neighbour Mn--Mn interactions in plane
(described by $J_a$) and out of plane (described by $J_c$).
Fitting our first-principles results to
this simple model we obtain the values for those constants that are displayed
in Fig.~\ref{fig_magnetism}(a) (bottom panels); the energies given by the 
model are represented by lines in the top panel, and they show reasonable
agreement with the first-principle data.
The exchange coupling constants for BiMnO$_3$ are very different from those
of BiFeO$_3$,\cite{Escorihuela2012PRL} reflecting the different natures
of Fe$^{+3}$ (with a $d^5$ electronic configuration of half-filled orbitals, 
leading to strong antiferromagnetic interactions according to the 
Goodenough-Kanamori rules of 
superexchange\cite{Goodenough1958JPCS,Kanamori1959JPCS}) and Mn$^{+3}$
(with a $d^4$ configuration that includes an empty $e_g$ orbital, favoring 
in-plane ferromagnetism---the Mn--O--Mn angles in
the $T_{aac}$ and $R_{aac}$ phases are between 150$^\circ$ and 160$^\circ$).
We then used a Monte Carlo method to solve our Heisenberg model in a 
periodically-repeated box
with $20 \times 20 \times 20$ spins; the results obtained for the order
parameter that describes the A-AFM alignment and for the magnetic 
susceptibility are shown in Fig.~\ref{fig_magnetism}(b).
Bulk BiMnO$_3$ orders magnetically below around
105~K,\cite{Bokov1966SPSS, Sugawara1968JPSJ} and the N\'eel temperatures of
our simulated films
are similarly low---around 80~K for the $T_{aac}$
film with lattice parameter $a_{\rm in} = 3.70$~\AA, and around
90~K for the $R_{aac}$ film with $a_{\rm in} = 4.08$~\AA.
The exchange coupling constants do not change much in the ranges of
epitaxial strains where the $T_{aac}$ and $R_{aac}$ are expected to be stable,
so the N\'eel temperature will be similar in these ranges.

\subsection{Optical Properties}

\begin{figure}
\centering
\subfigure[]{
\includegraphics[width=75mm]{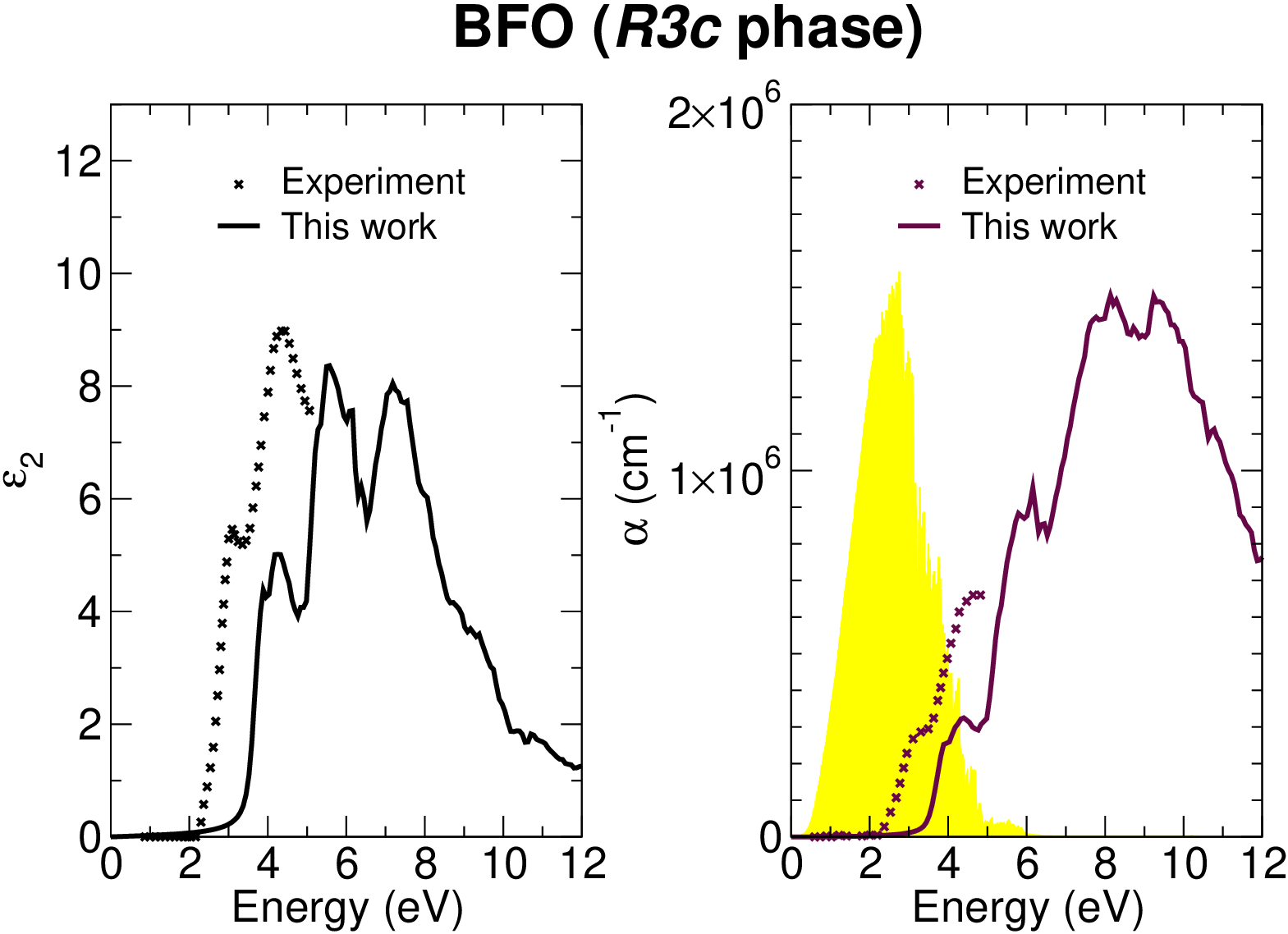}
}
\\
\subfigure[]{
\includegraphics[width=75mm]{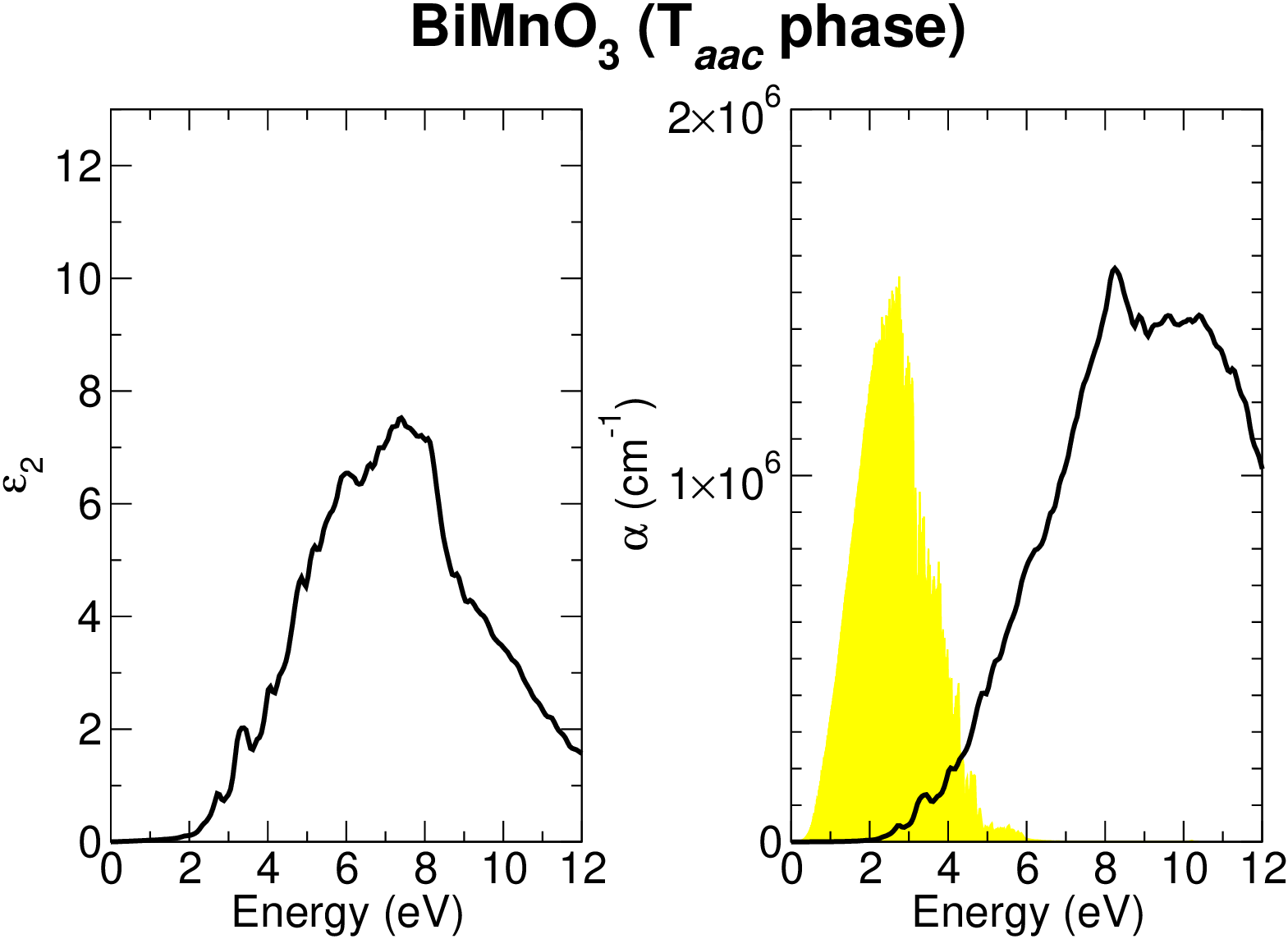}
}
\caption{(Color online.)
Imaginary component of the average of the diagonal elements of the dielectric
matrix (left) and absorption coefficient (right) for (a) bulk BiFeO$_3$, and
(b) a $T_{aac}$ film of BiMnO$_3$ ($a_{\rm in} = 3.70$~\AA).
The experimental values for BiFeO$_3$ were taken from
Ref.~\onlinecite{Chen2010APL}.
The shaded area corresponds to the solar spectrum.
}
\label{fig_optical}
\end{figure}

BiMnO$_3$ is a particular perovskite oxide not only in its ferromagnetic
properties, but also in its small band gap---typically, band gaps are
above 3~eV for these materials,\cite{Grinberg2013Nat} but they are around 2~eV
or less
for some of the phases of BiMnO$_3$ according to our calculations and
to the few experiments available.\cite{McLeod2010PRB}
In order to further investigate the optical properties of BiMnO$_3$, we 
used the independent particle approximation implemented
in {\sc Vasp}\cite{Gajdos2006PRB}
to compute the frequency-dependent dielectric matrix, and the related 
absorption coefficient.
Since no experimental data for comparison are available for BiMnO$_3$, we
did a initial test on BiFeO$_3$---we compared the absorption coefficient
of bulk BiFeO$_3$ computed following this methodology with the one measured
by Chen {\em et al.} for a film with a very similar 
structure.\cite{Chen2010APL}
Our BiFeO$_{3}$ results in Fig.~\ref{fig_optical}(a) show good 
agreement between theory and experiment, if we correct a shift 
associated to the
overestimation of the band gap by the theoretical method. 
(This band gap for BiFeO$_{3}$ (3.4~eV) is the same 
reported earlier by Stroppa and Picozzi,\cite{Stroppa2010PCCP} who
performed a calculation similar to ours.)
For the BiMnO$_3$ $T_{aac}$ film expected to be stable when grown on a substrate
of around 3.70~\AA, we find an absorption spectra that matches the solar range
better than what is typical in other perovskite oxides, as shown in
Fig.~\ref{fig_optical}(b).
We found a band gap of around 2~eV, but this could be even smaller if the HSE06
hybrid method is again overestimating it.
This makes supertetragonal BiMnO$_3$ films interesting in the framework
of materials for photovoltaic devices where light absorption can be
coupled to other functional properties.

%%%%%%%%%%%%%%%%%%%%%%%%%%%%%%%%%%%%%%%%%%%%%%%%%%%%%%%%%%%%%%%%%%%%%%%%%%%%%%%%

\section{Conclusions}

In this article we have explored the epitaxial phase diagram of BiMnO$_3$ 
films with the help of first-principles calculations.
While bulk BiMnO$_3$ is a paraelectric, we predict that it will transform
to a supertetragonal phase with a polarization of around 100~$\mu$C/cm$^2$
when grown on substrates that compress its in-plane lattice constant to about
3.75~\AA~(for example, YAlO$_3$ or LaSrAlO$_4$\cite{Schlom2007ARMR}).
This polarization will point mostly out of plane, but it is also possible
for BiMnO$_3$ films to develop a polarization of similar size laying mostly
in plane by growing it at tensile strains on top of substrates that expand 
its lattice constant beyond 4~\AA\ (such as BaTiO$_3$ or
PZT\cite{Schlom2007ARMR}).
Our findings might explain the experimental reports of ferroelectricity
in BiMnO$_3$ films grown on SrTiO$_3$ and LaAlO$_3$ with relatively large
remnant polarization;\cite{Jeen2011JAP, Son2008APL, DeLuca2013APL}
even if SrTiO$_3$ and LaAlO$_3$ have a larger lattice constant than the one
needed for stabilizing the supertetragonal phase according to our calculations,
these films show in experiments nonuniform strain
distributions,\cite{Jeen2011JAP} 
which could be the signature of a co-existence of different (polar and
non-polar) phases of BiMnO$_3$ films.

The supertetragonal phase that we have found is very similar to that
of BiFeO$_3$ films grown on LaAlO$_3$\cite{Bea2009PRL} 
($a_{\rm in} \approx 3.79$~\AA)
and of BiCoO$_3$, where it is the ground state even in bulk 
($a_{\rm in} \approx 3.73$~\AA\cite{Belik2006CM}).
This kind of phases with giant polarization might be ubiquitous
in transition-metal perovskite oxides containing bismuth, whose lone electron
pair can be easily accommodated in this kind of
structures;\cite{Dieguez2011PRB}
note that previous calculations for BiScO$_3$\cite{Iniguez2003PRB} 
provide a hint that this might be yet another material displaying the same
behaviour.
Since BiMnO$_3$ is the only strong ferromagnet among the
insulating transition-metal
perovskite oxides, our results are also relevant in the context of discovery
of new multiferroics; we have found that A-AFM ordering is energetically
favored, but the FM ordering is competitive (especially for the
$R_{aac}$ phase).
It might thus be possible to engineer ferroelectric ferromagnets by combining
these new film phases of BiMnO$_3$ with other transition-metals oxides 
in superlattices or solid solutions.
Finally, in addition to this functional properties, some of these phases
of BiMnO$_3$ show band gaps that are smaller than those found typically
in perovskite oxides, which makes this material interesting also from the
point of view of photovoltaic applications.

%%%%%%%%%%%%%%%%%%%%%%%%%%%%%%%%%%%%%%%%%%%%%%%%%%%%%%%%%%%%%%%%%%%%%%%%%%%%%%%%

\section*{Acknowledgements}
{O.D. acknowledges funding from the Israel Science Foundation through
Grants 1814/14 and 2143/14.
J.\'I. was financially supported by the Fond National de Recherche Luxembourg
through a PEARL grant (FNR/P12/4853155/Kreisel) and by MINECO-Spain
(Grant No. MAT2013-40581-P).
}

%%%%%%%%%%%%%%%%%%%%%%%%%%%%%%%%%%%%%%%%%%%%%%%%%%%%%%%%%%%%%%%%%%%%%%%%%%%%%%%%

%%%%%%%%%%%%%%%%%%%%%%%%%%%%%%%%%%%%%%%%%%%%%%%%%%%%%%%%%%%%%%%%%%%%%%%%%%%%%%%%

\end{document}